\documentclass[12pt]{iopart}
\usepackage{xcolor}
\usepackage{graphicx}

\DeclareMathAlphabet{\mathpzc}{OT1}{pzc}{m}{it}
\begin{document}

\title[Efimov effect in a $D$-dimensional Born-Oppenheimer approach]{Efimov effect in a $D$-dimensional Born-Oppenheimer approach}

%

\author{D S Rosa$^{1}$, T Frederico$^{2}$, G Krein$^{1}$ and M T Yamashita$^{1}$}
\address{$^{1}$Instituto de F\'\i sica Te\'orica, Universidade Estadual Paulista, Rua Dr. Bento 
Teobaldo Ferraz, 271 - Bloco II, 01140-070, S\~ao Paulo, SP, Brazil}
\address{$^{2}$Instituto Tecnol\'{o}gico de Aeron\'autica, 12228-900, S\~ao Jos\'e dos Campos, SP, Brazil}

\vspace{10pt}
\begin{indented}
\item[]August 2018
\end{indented}

\begin{abstract}
We study a three-body system, formed by two identical heavy bosons and a light 
particle, in the Born-Oppenheimer approximation for an arbitrary dimension $D$. We restrict 
$D$ to the interval $2\,<\,D\,<\,4$, and derive the heavy-heavy $D$-dimensional effective potential 
proportional to $1/R^2$ ($R$ is the relative distance between  the heavy particles), which is 
responsible for the Efimov effect. We found that the Efimov states disappear once the 
critical strength of the heavy-heavy effective potential $1/R^2$ approaches the limit 
$-(D-2)^2/4$. We obtained the scaling function for the $^{133}$Cs-$^{133}$Cs-$^6$Li system 
as the limit cycle of the correlation between the energies 
of two consecutive Efimov states as a function of $D$
and the heavy-light binding energy $E^{D}_2$. 
In addition, we found that the energy of the $(N+1)^{\rm th}$ excited state reaches the 
two-body continuum independently of the dimension $D$ when $\sqrt{E^{D}_2/E_3^{(N)}}=0.89$, 
where $E_3^{(N)}$ is the $N^{\rm th}$ excited three-body binding energy.
\end{abstract}

%
%
%
%
%

\section{Introduction}

In the Landau and Lifshitz book on nonrelativistic quantum mechanics~\cite{landau}, 
one can read: ``to reveal certain properties of quantum-mechanical motion it is useful to examine 
a case which, it is true, has no direct physical meaning: the motion of a particle in a field where 
the potential energy becomes infinite at some point (the origin) according to the law 
$U(r)\sim-\beta/r^2\,\,(\beta>0)$''. This book was first published in English language
in 1958 and the transcribed sentence appears at the beginning of 
the subsection ``fall of a particle to the centre'', in which the possible solutions of the 
Schr\"odinger equation for such a  potential are studied. In particular, its was
shown that the radial equation admits solutions of the form $r^{r_0+is}$, with $s$ being real for 
a range of values of the strength of the potential. In such a case, there are 
infinitely many bound states for the system with a spectrum unbounded from below. We 
recall that  the ``fall to the centre'' phenomenon was found much 
earlier by Thomas in 1935 \cite{thomas} for the triton in the limit of zero-range 
interaction between the nucleons. Later on, in 1970, Efimov found that a three-boson system 
presents an infinite number of three-body bound states for zero angular 
momentum \cite{efimov} in the limit of infinite two-body scattering lengths \textemdash in 
this case an attractive long range potential proportional to $ - 1/\rho^2$, where 
$\rho$ is the hyper-radial coordinate, appears with a strength large enough to collapse 
the system analogously to the Landau example. The counterintuitive phenomenon discovered by Efimov 
appeared to be only a theoretical speculation added to the fact that the study was originally made 
in the nuclear physics context, where there is no possibility to have a two-body zero binding energy.

After almost thirty years since Efimov's discovery, the experimental group from Innsbruck 
finally observed indirectly the formation of Efimov molecules in ultracold atomic 
traps \cite{grimm} by using the Feshbach resonance phenomenon~\cite{chin} to freely tune 
the two-body scattering lengths. The observation was made through the measurement of the 
three-body atomic loss peaks. These peaks appear as a consequence of the resonant three-body 
recombination process, where three atoms recombine forming a deep bound pair plus an atom with a 
recoil energy larger than the energy of the trap, resulting in the loss of three atoms 
each time this process happens. The positions where these peaks appear are given by 
the two-body scattering lengths, $a_-^{(N)}$, the subindex 
refers to negative scattering lengths, meaning that there are no shallow two-body bound states,
and $N=0,1,2,\dots$ refers respectively to the ground, first, second, $\dots$ excited states. 
The ratio of consecutive $a_-^{(N)}$ follows the 
discrete Efimov scaling factor, $s$, and is given by 
$a_-^{(N+1)}/a_-^{(N)}={\rm exp}(\pi/s)$ \cite{ulmanis,mathias}. Thus, the observation 
of the position of the recombination peaks allows an indirect verification of the 
Efimov phenomenon. Nowadays, very advanced techniques allow the direct measurement of 
the binding energy of the three- and two-body molecules in atomic 
clouds~\cite{kunitski,zeller}. 

Fonseca and collaborators showed in a seminal paper \cite{fonseca} that the similarity 
between the Landau's ``fall to the centre'' and the general picture given by Efimov 
could be extended to strongly mass asymmetric systems, where the Born-Oppenheimer 
approximation applies. The authors of Ref.~\cite{fonseca} 
have shown that the exchange of the light particle between the heavy ones generates 
an effective potential proportional to $-1/R^2$, where $R$ is the separation distance between the 
heavy particles, which causes the accumulation of the three-body levels close to the continuum. 
Not only the form of the potential matters to have infinitely many three-body bound states but 
also the strength, which is directly related to the Efimov discrete scaling factor. 
In three-spatial dimensions, the form of the potential is presently well-known  
to be $-(s^{2}+1/4)/R^2$, and Efimov states with the characteristic 
log-periodic behavior will be present when $s^{2} > 0$.

Nowadays, there is an increasing interest in studying weakly bound 
few-body systems in spatial dimension different than three \cite{moller}.  
Experimentally, dimensional transition can be obtained in 
ultracold atomic clouds in a confining potential squeezed 
asymmetrically in one or two directions. Consequentely, the 
atomic clouds can be compressed, or expanded, in such a way the 
effective-spatial dimension felt by the system is continuously 
changed \cite{pethick}.

Nielsen and collaborators \cite{nielsen} showed that for three identical bosons the Efimov 
effect exists in the interval $2.3\,<\,D\,<\,3.8$, recently this result has been proven using a 
different approach \cite{mohapatra}. More recently, we have extended 
this result to a heteroatomic $AAB$ system \cite{derick} based on the Danilov's 
solution \cite{danilov} of the  Skornyakov  and  Ter-Martirosyan three-boson integral
equation~\cite{skt} in the ultraviolet region. In Ref.~\cite{derick} 
we have determined the dimensional boundaries as a function of the atomic mass ratio, 
$\mathpzc{A}=m_B/m_A$, for which the Efimov effect exists, and the corresponding
Efimov discrete scaling factor was obtained as a function
of $\mathpzc{A}$ and $D$.

In the present work we study, in $D$ dimensions, a three-body $AAB$ system formed by 
two-identical heavy atoms $A$ and a different one $B$ in the Born-Oppenheimer approximation. 
This work extends our previous study in Ref.~\cite{derick} in several directions. 
First, we depart from the ideal case without scales studied in that reference. We generalize the 
work by Fonseca, Redish and Shanley~\cite{fonseca} to arbitrary $D$ spatial dimensions and mass 
ratios $\mathpzc{A}=m_B/m_A$, and calculate the effective heavy-heavy potential responsible for the 
appearance of the Efimov effect. We also obtain the $D$ and $\mathpzc{A}$ dependence of the 
critical strength for the existence of Efimov states, which for $D=3$ was determined by Landau 
and Lifshitz~\cite{landau} to be $-1/4$.  For completeness, we have also show that the new results 
agree in the appropriate limits with those in Ref.~\cite{derick}. In addition, we obtain
the full three-body spectrum. The spectrum reveals a new and very counterintuitive result, 
in that the ratio of the two-body binding energy $E^{D}_2$ to the $N-$th excited three-body energy 
$E_3^{(N)}$ for which the Efimov effect disappears is independent of $D$.

The article is organized as follows. In Section II we describe our formalism: 
the effective potential, coming from the light-atom equation, is extracted in momentum space for a 
zero-range interaction between the light-heavy atoms. The small distance regime of the effective potential 
gives the critical strengths for which the Efimov effect exists. In Section III we present the Efimov discrete 
scaling factor for a general value of $D$ and calculate the three-body energy 
spectrum using a $D$-dimensional Schr\"odinger equation. In section IV we summarize and conclude.

\begin{figure}[htb!]
\begin{center}
\includegraphics[width=8.5cm]{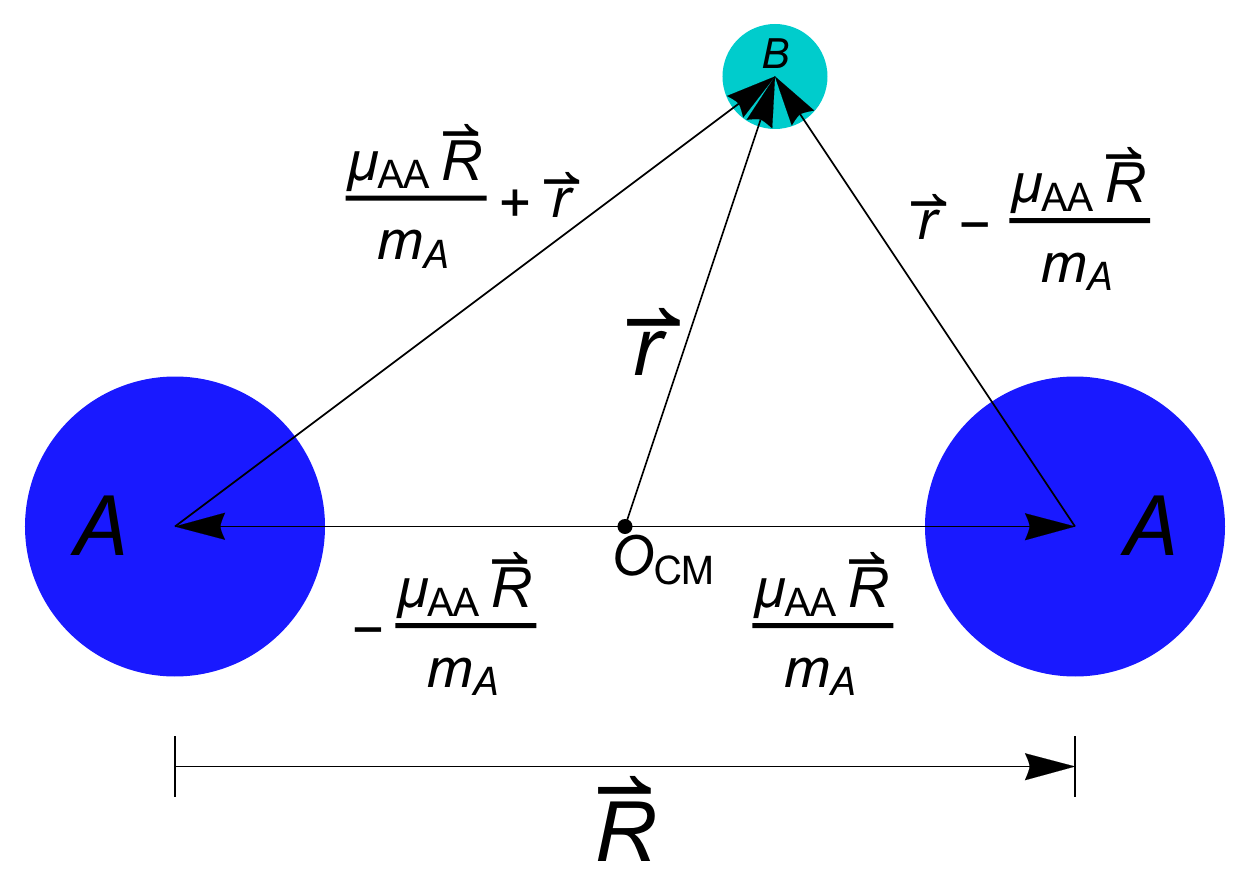}
\end{center}
\caption{ Three-body system formed by two identical heavy bosons with mass
$m_A$ and an particle with mass $m_B$. We are 
considering the Born-Oppenheimer validity range where $m_B/
m_A \ll 1$. }
\label{figsystem}
\end{figure}

\section{The Born-Oppenheimer approximation}

We consider a system formed by two-heavy bosons and a light particle,
pictorially represented in Fig. \ref{figsystem}. We employ the Born-Oppenheimer 
approximation; which allows us to separate the full three-body
Schr\"odinger equation into two equations, one for the heavy-light subsystem

\begin{eqnarray}
\fl \left[ -\frac{\hbar^{2}}{2\mu_{B,AA}}\nabla^{2}_{r} 
+ V_{AB}(\vec{r}-\frac{\mu_{AA}}{m_{A}}\vec{R})  
+ V_{AB}(\vec{r}+\frac{\mu_{AA}}{m_{A}}\vec{R})\right]\psi_R(\vec{r}) 
= \epsilon(R) \psi_R(\vec{r}), 
\label{light1}
\end{eqnarray}

\noindent and another for the two heavy particles:
\begin{equation}
\left[-\frac{\hbar^{2}}{2\mu_{AA}}\nabla^{2}_{R}  + V_{AA}(\vec{R}) + \epsilon(R) \right]\phi(\vec{R}) = 
E_3 \phi(\vec{R}),
\label{heavy}
\end{equation}
where the reduced masses are given by $\mu_{AA}=m_A/2$ and 
$\mu_{B,AA}=2m_A m_B/\left(2m_A+m_B\right)$, and $V_{AB}$ and $V_{AA}$ denote in an obvious 
notation the $AB$ and $AA$ two-body interactions, respectively. As usual with the Born-Oppeheimer 
approximation, $R$ enters as parameter in Eq.~(\ref{light1}) and can be used as a labelling index, 
and the eigenvalue $\epsilon(R)$ of the heavy-light equation enters as an effective potential in 
Eq.~(\ref{heavy}) for the heavy-heavy system. We employ a contact interaction with strength 
$\lambda$ for the heavy-light potential $V_{AB}$ so that one can write Eq.~(\ref{light1}) in 
momentum space as
\begin{equation}
\frac{q^2}{2\mu_{B,AA}} \tilde{\psi}_R(\vec{q}) 
+ \lambda \left[ e^{i \vec{q}\cdot\vec{R}/2} \, A_R^{(+)}
+ e^{-i\vec{q}\cdot\vec{R}/2} \, A_R^{(-)}\right] 
= \epsilon(R) \tilde{\psi}_R(\vec{q}),
\label{mom}
\end{equation}

\noindent where
\begin{equation}
A_R^{(\pm)}=\int \frac{d^Dq}{(2\pi)^D} \, 
e^{ \mp i \vec{q}\cdot\vec{R}/2 } \, \tilde{\psi}_R(\vec{q}),
\label{Apm_R}
\end{equation}
with $\tilde{\psi}_R(\vec{q})$ being the Fourier transform of $\psi_R(\vec{r})$, defined as
\begin{equation}
\psi_R(\vec{r}) = \int \frac{d^Dq}{(2\pi)^{D/2}} \, e^{i \vec{q}\cdot\vec{r}} \, \tilde{\psi}_R(\vec{q}) .
\end{equation}

The eigenvalue $\epsilon(R)$ can be determined as follows~\cite{bellotti}. 
Initially, one rewrites Eq.~(\ref{mom}) as
\begin{equation}
\tilde{\psi}_R(\vec{q}) = \lambda \,  
\frac{e^{i \vec{q}\cdot\vec{R}/2}  A_R^{(+)}+ e^{-i\vec{q}\cdot\vec{R}/2} \, A_R^{(-)}}{\epsilon(R) - \frac{q^2}{2\mu_{B,AA}}} .
\end{equation}
Next, one eliminates $\tilde{\psi}_R(\vec{q})$ in favor of $A_R^{(\pm)}$ using Eq.~(\ref{Apm_R}):
\begin{equation}
A_R^{(\pm)} = \lambda\int \frac{d^Dq}{(2\pi)^D} \; \frac{A_R^{(\pm)}
+ e^{\mp i\vec{q}\cdot\vec{R}} \, A_R^{(\mp)}}{\epsilon(R) - \frac{q^2}{2\mu_{B,AA}}} .
\end{equation}
Then, it is easily shown that for nontrivial solutions,  $A_R^{(\pm)}\neq 0$, the eigenvalue
is given by the transcendental equation:
\begin{equation}
\label{trans}
\frac{1}{\lambda} = \int \frac{d^Dq}{(2\pi)^D}
\, \frac{\cos(\vec{q}\cdot\vec{R}) + 1 }{\epsilon(R)-\frac{q^2}{2\mu_{B,AA}}}.
\end{equation}
The integral is divergent but the divergence can be dealt with by eliminating the strength 
$\lambda$ in favor of the two-body binding energy. That is, assuming that the two-body subsystem
contains a bound state with energy $E_{AB}\equiv-|E^{D}_2|$
\begin{equation}
\label{bound}
\frac{1}{\lambda} = - \int \frac{d^Dq}{(2\pi)^D}
\, \frac{ 1 }{|E^{D}_2| + \frac{q^2}{2\mu_{B,AA}}},
\end{equation}
and using this to replace $1/\lambda$ in Eq. (\ref{trans}) one obtains
\begin{equation}
\int \frac{d^Dq}{(2\pi)^D} \left[
\frac{\cos(\vec{q}\cdot\vec{R})+1}{\epsilon(R)-\frac{q^2}{2\mu_{B,AA}}}
+\frac{1}{|E^{D}_2|+\frac{q^2}{2\mu_{B,AA}}} \right] = 0.
\label{effpot}
\end{equation}

The integral in Eq. (\ref{effpot}) can be solved analytically; the
result is
\begin{eqnarray}
2^{D/2}{\bar R}^{1-D/2}|\bar{\epsilon}({\bar R})|^{(D-2)/4}
K(D/2-1,{\bar R}|\bar{\epsilon}({\bar R})|^{1/2})
&& \nonumber \\ 
-  \pi\csc\left(D\pi/2\right) 
\frac{1}{\Gamma(D/2)} \,
\left(1-|\bar{\epsilon}
({\bar R})|^{D/2-1}\right) = 0&&, \;\;\;\;\;
\label{fullequation}
\end{eqnarray}
where $|\bar{\epsilon}(R)| = |\epsilon(R)|/|E^{D}_{2}|$ and ${\bar R}=R/a$ 
with $a=\sqrt{\hbar^{2}/{2\mu_{B,AA}|E^{D}_{2}|}}$. $K(z)$ and $\Gamma(z)$ are, respectively, 
the modified Bessel function of the second kind and the gamma function.

The  effective potential $\bar{\epsilon}(R)$ is obtained from 
the transcendental equation in Eq.~(\ref{effpot}) that can be solved for 
a given mass ratio $\mathpzc{A} = m_B/m_A$ and dimension $D$. The 
effective potential assumes quite simple forms in the limits of large and small 
${\bar R}$. In the following, we  investigate both limits and compare 
them with the well known results for $D=3$ and $D=2$.

\vspace{0.25cm}
\noindent
{\it Large distances regime, ${\bar R}\gg1$}. For large ${\bar R}$ the light atom bounds to only one of 
the heavy atoms in such a way the three body problem is roughly reduced to a two body problem with 
$|E_{3}|\rightarrow|E^{D}_{2}|$. The effective potential can then be written as 
$\lim_{{\bar R}\rightarrow\infty}|\epsilon({\bar R})| = |E^{D}_{2}|+V({\bar R})$, with $V({\bar R})\rightarrow 0$ for 
${\bar R} \rightarrow \infty$. Replacing this asymptotic result in Eq. (\ref{fullequation}) we have that the 
effective potential can be written as 
\begin{equation}
|\bar{\epsilon}_{_{{\bar R}\rightarrow \infty}}({\bar R})| = 1 
+ \frac{K\left(D/2-1,{\bar R}\right)}{d(D,\bar{R})} ,
\end{equation}
where 
\begin{eqnarray}
d(D,\bar{R}) &=& \frac{{\bar R} }{2}K(D/2,{\bar R})
+ (1-D/2) \nonumber \\
&&\times \Biggl[ \frac{ \pi \csc(D\pi/2)}{2^{D/2}{\bar R}^{(1-D/2)}\Gamma(D/2)} \nonumber \\
&&  +\,  K(D/2-1,{\bar R}) 
\Biggr].
\end{eqnarray}
For $D=3$, one obtains well-known result of Ref.~\cite{fonseca}:
\begin{equation}
|\bar{\epsilon}_{_{{\bar R}\rightarrow \infty}}({\bar R})|^{D=3} = 1 + 
\frac{2}{{\bar R}} e^{-{\bar R}} ,
\end{equation}
and for $D=2$, the result of Ref.~\cite{bellotti}:
\begin{equation}
|\bar{\epsilon}_{_{{\bar R}\rightarrow \infty}}({\bar R})|^{D=2} = 1+ \frac{2K(0,{\bar R})}{1+{\bar R} K(1,{\bar R})}.
\end{equation}

\vspace{0.25cm}
\noindent
{\it Small distances regime, ${\bar R}\rightarrow0$}. This regime is directly related to the 
appearance of the Efimov effect. For $D=2$ the potential presents a Coulombic behaviour 
reproducing exactly a previous result obtained in Ref.~\cite{derickBO}, namely:
\begin{equation}
|\bar{\epsilon}_{_{{\bar R}\rightarrow 0}}({\bar R})|^{D=2} = \frac{2}{{\bar R}} e^{-\gamma} ,
\end{equation}
where $\gamma = 0.57721 \cdots $ is the Euler-Mascheroni number.
For values of $D$ in the interval $2<D<4$, it is convenient rewrite Eq. (\ref{fullequation}) as
\begin{equation}
\frac{|\bar{\epsilon}(\bar{R})|^{D/4-1/2}}
{1-|\bar{\epsilon}(\bar{R})|^{D/2-1}} = 
\frac{ \pi \csc (\pi D/2) \bar{R}^{D/2-1} }
{2^{D/2}\Gamma(D/2) K(D/2-1,\bar{R} {|\bar{\epsilon}(\bar{R})|^{1/2}})}.
\label{shorts}
\end{equation}
As the effective potential 
diverges at short distances, one can isolate $|\bar{\epsilon}(\bar{R})|$ in 
Eq. (\ref{shorts}) and write
\begin{equation}
|\bar{\epsilon}_{R \rightarrow 0}(\bar{R})| = \frac{g(D)}{\bar{R}^{2}} ,
\label{gamma}
 \end{equation}
where $g(D)$ is the solution of the transcendental equation
\begin{equation}
g(D) = \left[-\frac{ \pi \csc(D \pi/2)}
{ 2^{D/2} \Gamma({D}/{2}) K(D/2-1,\sqrt{g(D)}) } \right]^{\frac{4}{2-D}} .
\end{equation}
For $D=3$, one obtains
\begin{equation}
g(3) = 0.3216,
\end{equation}
which reproduces the $D=3$ results \cite{fonseca,mathias}. The effective strength $g(D)$ plays 
here a central role in the occurance of the Efimov effect for $D>2$. This will be discussed in 
the next section.

\section{Efimov effect for $D$ Spatial Dimensions}

\begin{figure}[htb!]
\centering
\includegraphics[width=9cm]{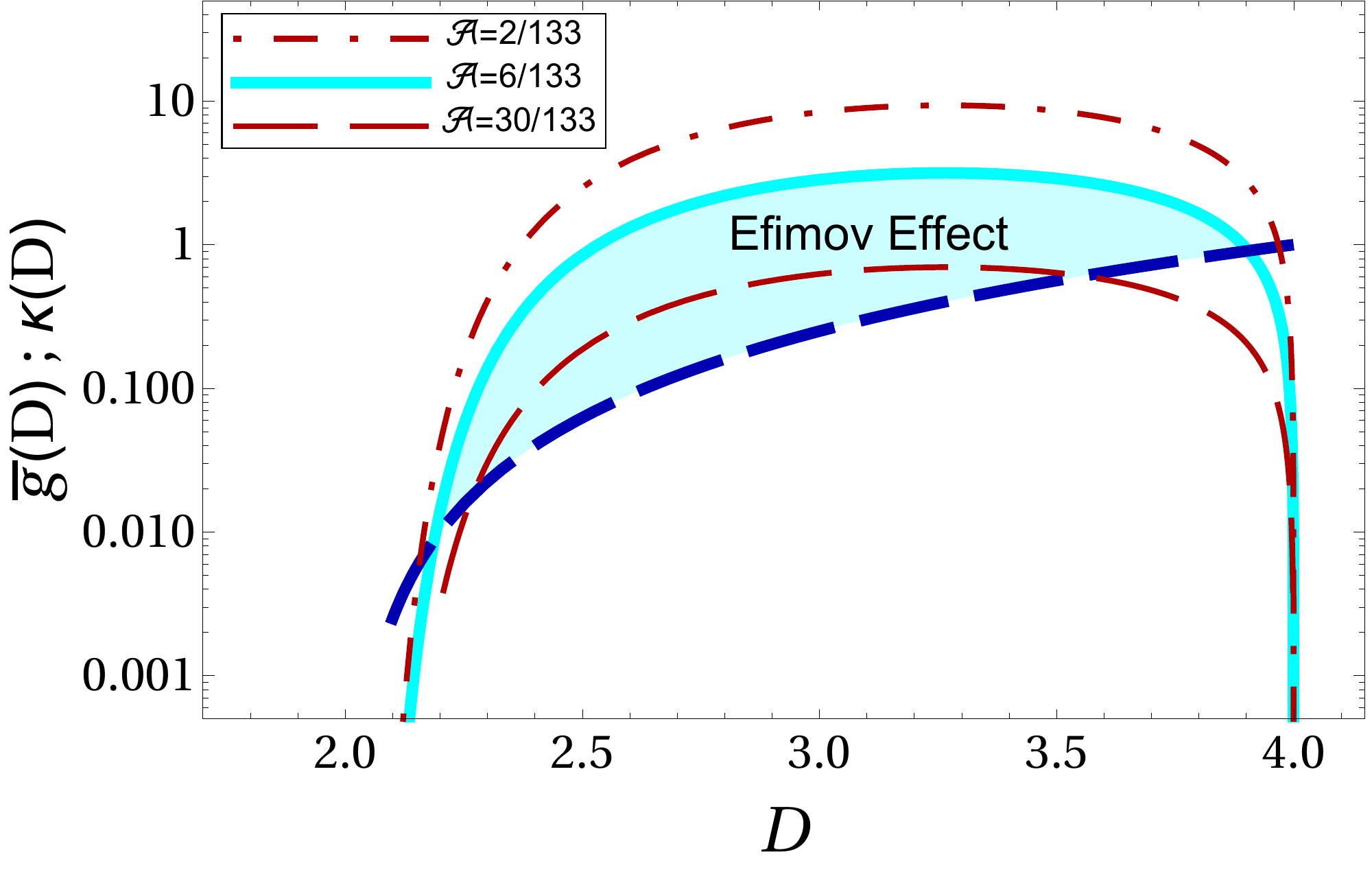}
\caption{The shadowed area shows the region where the Efimov effect is allowed according to 
the effective strength given by Eq. (\ref{gamma}). The blue dashed line is the critical 
strength given by $\kappa(D)=(D-2)^{2}/4$ Eq. (\ref{scalingD}), and the effective strength 
${\bar g}(D)=(\mathpzc{A}+2)g(D)/4\mathpzc{A}$ is presented for three 
different mass ratios. The dimensional 
limit is $2.21<D<3.90$ for $\mathpzc{A}=6/133$.} 
\label{figDlimits}
\end{figure}

The $D$-spherical Schr\"odinger equation \cite{martins} for zero total angular momentum is given by
\begin{equation}
\left[-\frac{\hbar^{2}}{2\mu_{AA}}\nabla^{2}_{R}  + \epsilon(R) \right]\phi(\textbf{R}) = E \phi(\textbf{R}),
\end{equation}
where the radial part of the Laplacian reads
\begin{equation}
\nabla^{2}_{R}  = \frac{d^{2}}{dR^{2}}+\frac{D-1}{R}\frac{d}{dR}.
\end{equation}
For two identical heavy particles and considering an infinitely high excited three-body state 
with energy $E_{3}\approx0$ we can write
\begin{equation}
\left[\frac{d^{2}}{dR^{2}}+\frac{D-1}{R}\frac{d}{dR}  - \frac{2\mu_{AA}}{\hbar^{2}} \epsilon(R) \right]\phi(R) = 0.
\end{equation}
The effective potential in the regions where the Efimov effect appears (small distances) is given by 
Eq. (\ref{gamma}), $-{\cal G}/R^{2}$, where ${\cal G}$ depends on the mass ratio and dimension. Replacing 
the asymptotic effective potential and using the ansatz $\phi(R)=R^{\delta}$ we can calculate the 
Efimov discrete scaling factor $s$ for a general $D$ $(2<D<4)$.
\begin{equation}
\delta= \frac{1}{2}\pm i \sqrt{\frac{\mathpzc{A}+2}{4 \mathpzc{A}}g(D) - (D/2-1)^{2}}=\frac{1}{2}\pm i s.
\label{scalingD}
\end{equation}
In order to have the Efimov effect, $\phi(R)$ should oscillate, thus $s$ should be real. 
As an example, Figure \ref{figDlimits} shows for a $^{133}$Cs-$^{133}$Cs-$^6$Li system, $\mathpzc{A}=6/133$, 
the region where the Efimov effect is allowed. The difference of the $D$ limits of the BO approximation 
with the exact result \cite{derick} is less than 2\%. Note that the effect of a finite $E^{D}_2$ is washed 
out here as the critical strength is obtained in the limit of ${\bar R}\rightarrow0$.

Our previous results generalizing the Efimov discrete scaling factor were 
obtained in the ideal unitary limit~\cite{derick}, where all the energies 
are washed out of the problem. As energies are always finite in experiments, 
in this section we generalize the Schr\"odinger equation to $D$ spatial 
dimensions to obtain the three-body energy spectrum:

\begin{equation}
\left[-\frac{d^{2}}{dR^{2}} - \frac{D-1}{R} \frac{d}{dR}  
+ \frac{m_A}{\hbar^2} \epsilon(R)  
+ \frac{l(l+D-2)}{R^2} \right] \phi(R) = \frac{m_A}{\hbar^2} E_3 \, \phi(R) .
\end{equation}
Replacing $\phi(R) = \chi(R)/R^{(D-1)/2}$, this equation becomes
\begin{equation}
\left[-\frac{d^{2}}{dR^{2}}  +\frac{f(D,l)}{R^{2}} 
+ \frac{m_A\epsilon(R)}{\hbar^{2}} \right]\chi(R) 
= \frac{m_AE_3}{\hbar^{2}}\chi(R),
\label{Sch-3}
\end{equation}
where $f(D,l) = [D^2-4D+3+4l(l+D-2)]/4$. Note that $f(D,l)$ reduces to 
the well-known results $f(2,l) = (-1+4l^2)/4$ and  $f(3,l) = l(l+1)$. 
Rescaling $R$ and $\epsilon(R)$ as in the previous section, Eq.~(\ref{Sch-3}) can be rewritten as
\begin{equation}
\left[- \frac{d^2}{d\bar{R}^2} + \frac{f(D,l)}{\bar{R}^{2}}
+ \frac{m_A}{2\mu_{B,AA}} \bar{\epsilon}(\bar{R})\right] \chi(\bar{R}) 
= \frac{m_A}{2\mu_{B,AA}}\frac{E_{3}}{|E^{D}_2|} \chi(\bar{R}).
\end{equation}

The potential diverges  for 
${\bar R} \rightarrow 0$ and needs to be regularized. We choose 
a regularization function of the form 
$1-e^{-(R/R_0)^3}$,  where the parameter $R_0$ is related to the van de Waals length that 
cuts off the very short distance region 
related to the chemistry of the heavy atoms. Higher powers of the regularization function can be compensated with a slight change 
of $R_0$ in order to preserve the present results.
The regularized Schr\"odinger
equation in $D$ dimensions is given by 
\begin{eqnarray}
\fl \left[-\frac{d^{2}}{d\bar{R}^{2}} + \frac{f(D,l)}{\bar{R}^{2}} 
+ \frac{m_A}{2\mu_{B,AA}} \left(1-e^{- (\bar{R}/\bar{R}_0)^3}\right) \bar{\epsilon}(\bar{R})\right] 
\chi(\bar{R}) 
= \frac{m_A}{2\mu_{B,AA}}\frac{E_{3}}{|E^{D}_2|}\chi(\bar{R}),
\label{schD}
\end{eqnarray}
with $\bar{R}_0 = R_0/a$, where $a$ was defined below Eq.~(\ref{fullequation}).

\begin{figure}[htb!]
\centering
\includegraphics[width=9.2cm]{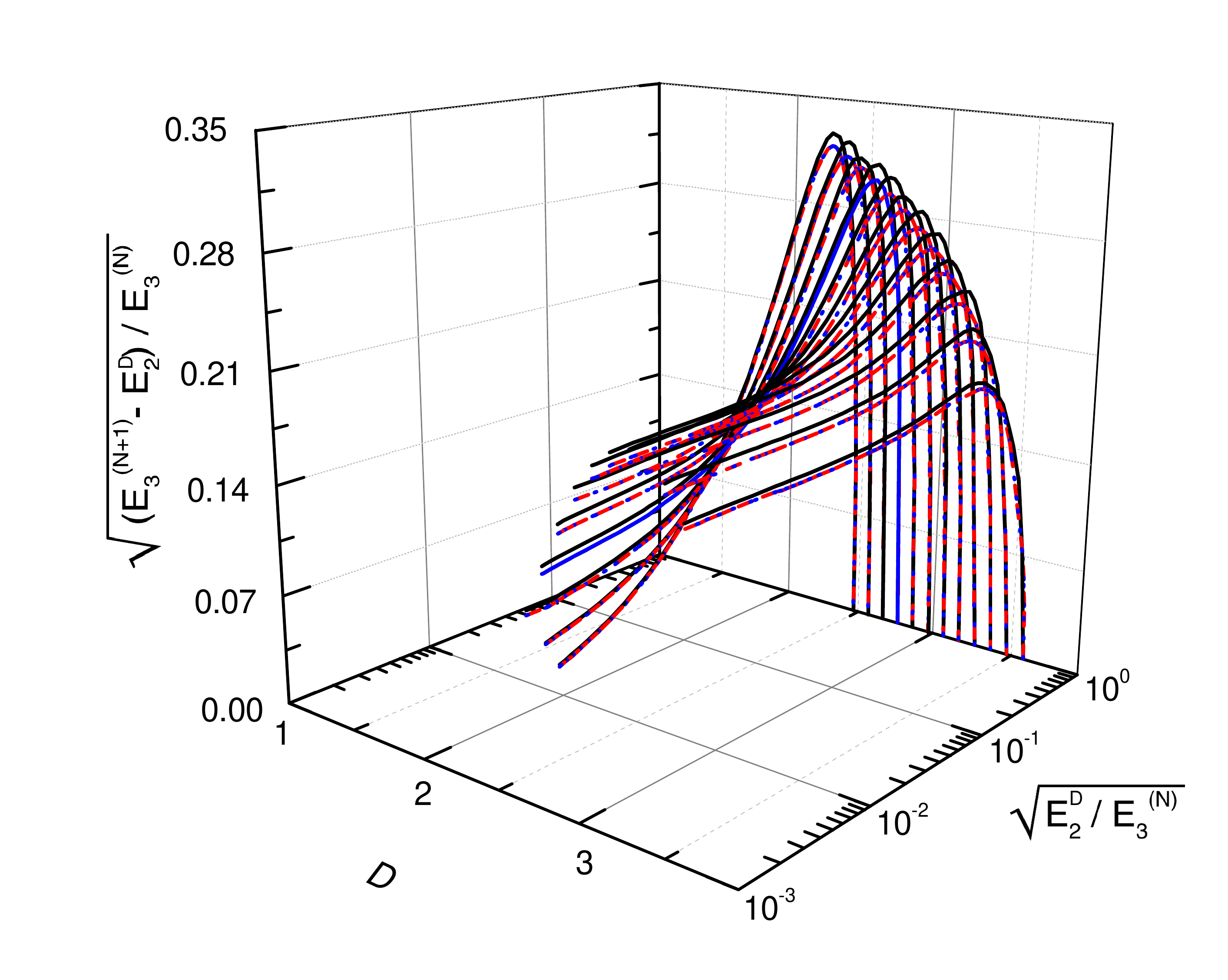}
\includegraphics[width=8.3cm]{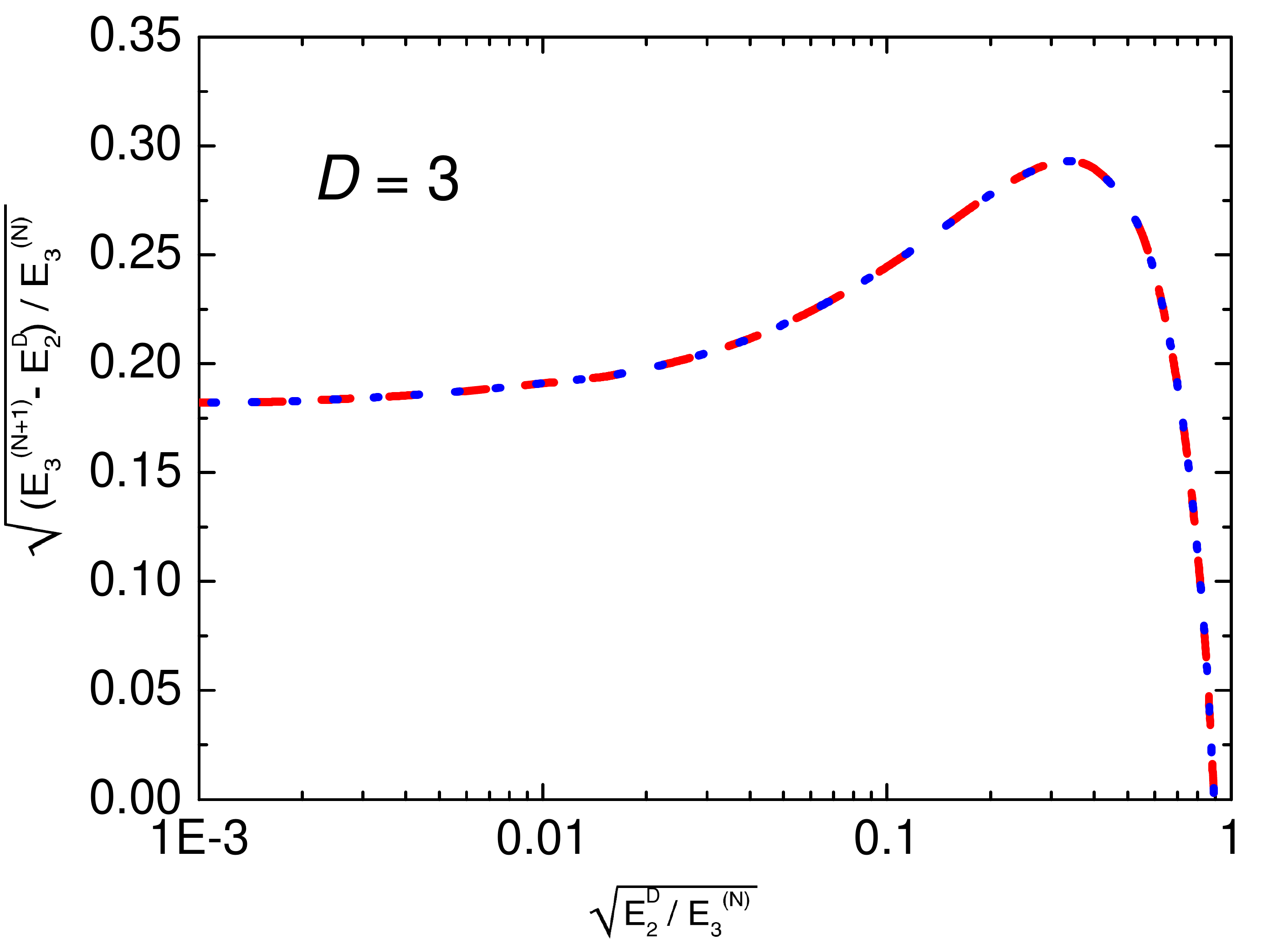}
\caption{Upper frame: three-body energies for the $^{133}$Cs-$^{133}$Cs-$^6$Li system, $\mathpzc{A}=6/133$, 
for different dimensions. The energies are obtained from the numerical solution of 
Eq. (\ref{schD}). The solid black, dashed red and dotted blues lines are, respectively, 
results for $N=0,\,1$ and 2. Lower frame: curve for $D=3$.}
\label{figenergy}
\end{figure}

The three-body energy calculated from Eq. (\ref{schD}) can be represented in terms 
of a universal scaling function for two consecutive  trimer energies
$E_3^{(N)}$ and $E_3^{(N+1)}$, where higher indexes indicate higher excited states (see e.g. \cite{tobias}), as
\begin{equation}
\lim_{N\rightarrow\infty}\sqrt{\frac{E_3^{(N+1)}-E^{D}_2}{E_3^{(N)}}}
={\cal F}\left(\sqrt{\frac{E^{D}_2}{E_3^{(N)}}},\,D,\, \mathpzc{A}\right).
\label{scaling}
\end{equation}
It is important to note that the limit in Eq. (\ref{scaling}) is 
reached very fast and defines the universal scaling function ${\cal F}$, obtained 
numerically, as showed in Fig. \ref{figenergy}. 

The upper frame of Fig. \ref{figenergy} was constructed using $N=0,\,1$ and 2, and the 
corresponding results are represented by the solid (black), dashed (red) and dotted (blue) 
lines, respectively. The coincidence between dashed and dotted lines 
shows that the limit cycle is already obtained for $N=1$. The curve for $D=3$ is showed in the 
lower frame of Fig. \ref{figenergy} where the constant three-body energy ratio becomes evident once the magnitude of $E^{D}_2/E_3^{(N)}$ 
decreases. The Efimov discrete scaling factor, $s$, can then be extracted using  
Eqs. (\ref{gamma}) and (\ref{scalingD}) or checking the three-body energy ratios for small 
$E^{D}_2$ - both methods give exactly the same result, which differs from the exact calculation by 
less than 2\% \cite{derick} for the boundaries of existence of the Efimov effect. 

It is important to note that the universal point where an excited three-body state 
disappears at the two-body energy cut does not depend on the dimension of the 
system, it depends only on the mass ratio. For $\mathpzc{A}=6/133$ the critical 
point is given by $\sqrt{E^{D}_2/E_3^{(N)}}=0.89$. Possible effects coming from a 
finite $E^{D}_2$ at the threshold for 
the disappearance of an excited state, observed in Ref. \cite{ulmanis}, are 
not increased by the dimensional change as the critical points do not depend on 
$D$.

\begin{figure}[htb!]
\centering
\includegraphics[width=9.5cm]{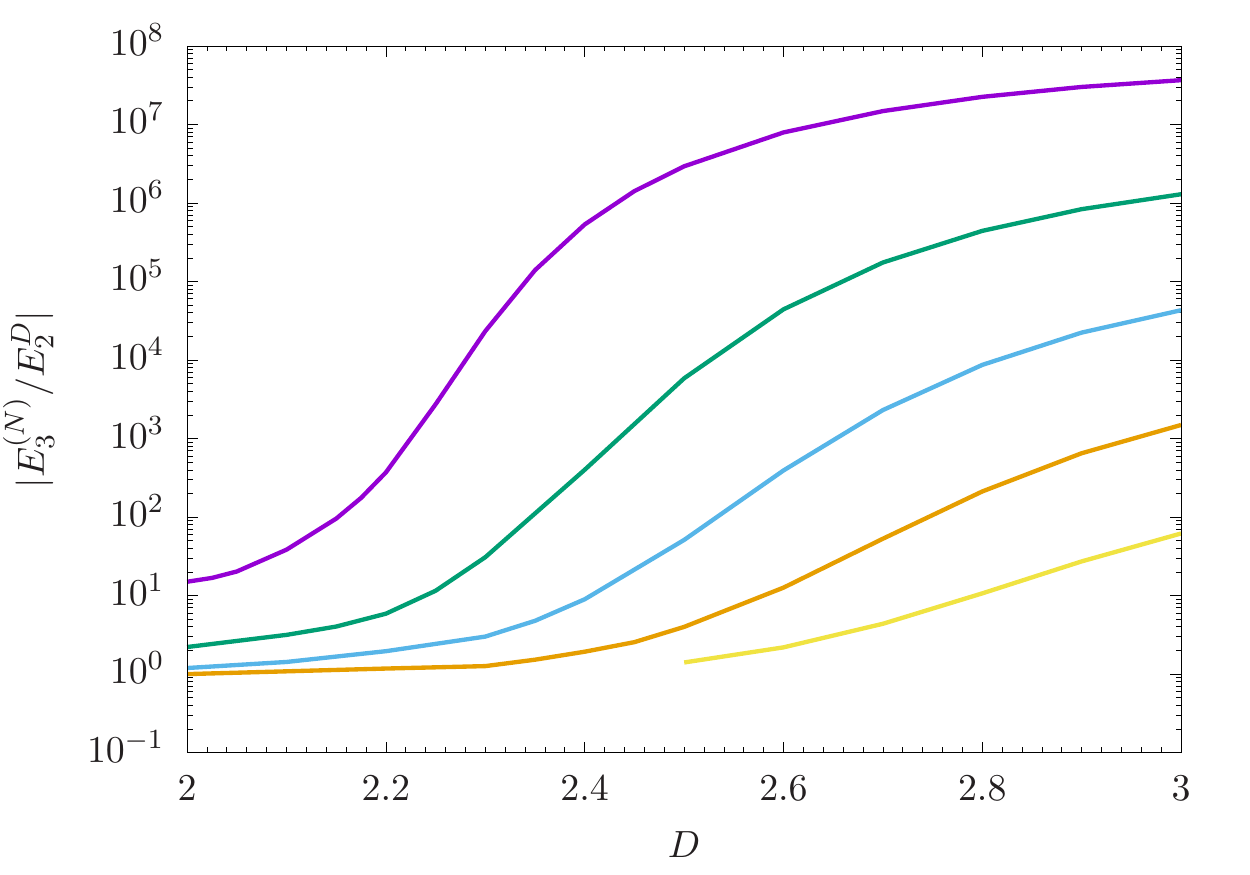}
\includegraphics[width=9.5cm]{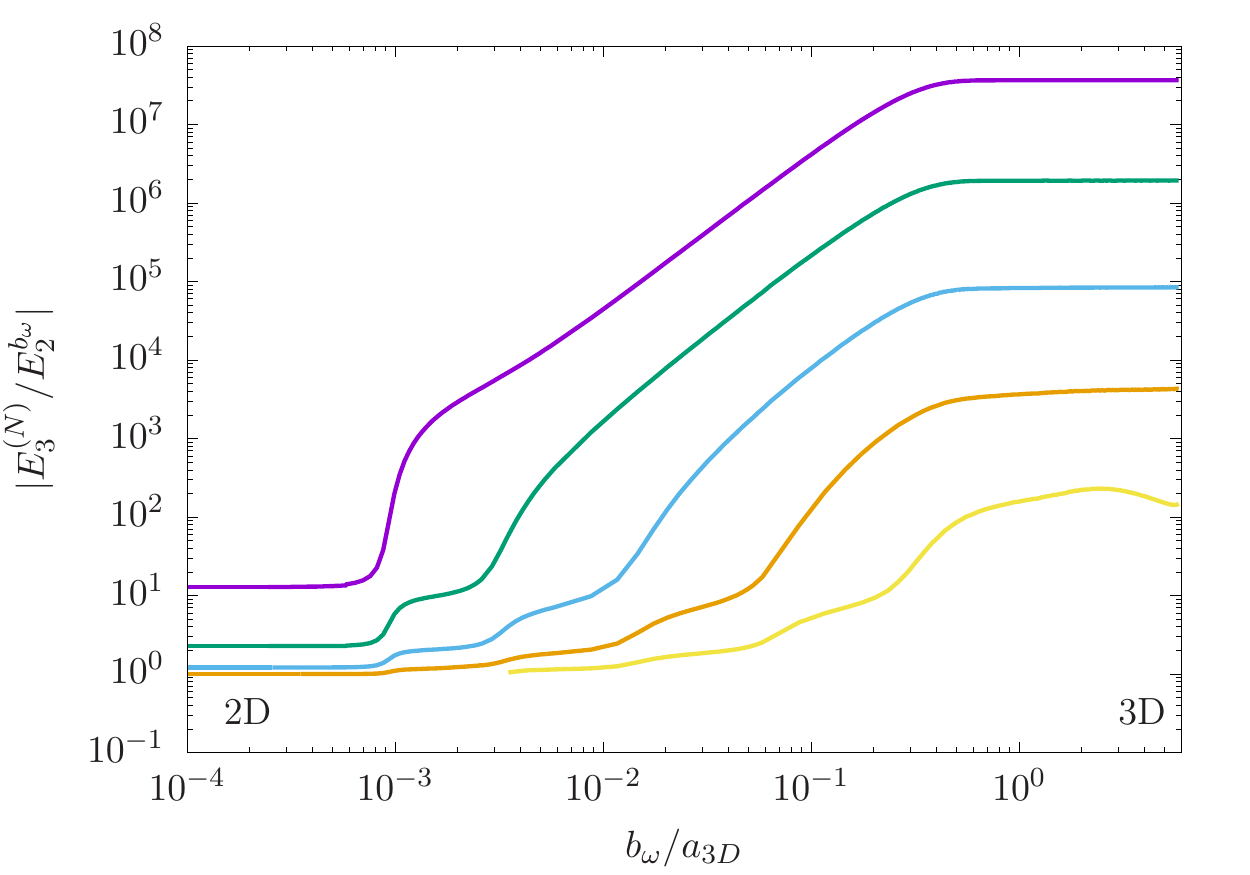}
\caption{Three-body energy spectrum normalized to the two-body energies in $D$ dimensions for 
$\mathpzc{A}=6/133$. Upper panel: spectrum calculated with the Schr\"odinger equation in 
Eq.~(\ref{schD}). Lower panel: spectrum calculated with a full Faddeev calculation~\cite{sandoval} 
in a cubic box, where $b_\omega$ is the equivalent squeezing parameter of a harmonic trap{\textemdash}see 
text for explanations.}
\label{figenergy3}
\end{figure}

The connection of the results presented in this article with the 
experiment can be made clear when we compare the energy spectrum of the three-body 
system found in this work with the three-body spectrum of a full 3D calculation in the 
presence of a external trap \cite{sandoval}. In Fig. \ref{figenergy3} we present in the upper frame the three-body 
energy spectrum for a $^{133}$Cs-$^{133}$Cs-$^6$Li system \cite{chin2014} for fractional dimension when the two-body 
energy depends on the dimension, in the lower frame the three-body energy spectrum is 
presented when the two-body energy depends on the squeezed parameter $b_{\omega}$ for the $^{133}$Cs-$^{133}$Cs-$^6$Li system. 
The results presents similar qualitative results mostly in the limit of two and three dimensions. 
The connection for fractional dimensions \cite{levinsenprx} is not clear once the relation of the dimension with the 
squeezed parameter of the trap is not clear and is not the aim of this work to prove such relation once 
this would demand an elaborate calculation for three particles in external fields implying complications 
as four-body problem \cite{esben}. On the other hand, when a three-body system in the presence of an external trap approaches 
two dimensions, the infinite bound states present in fractional dimensions need to fit into the four possible 
bound states in two dimensions, causing an avoided crossing effect that becomes quite sharp in Ref. \cite{sandoval} when 
the energy of three-body is calculated for fixed values of the two-body energy. However, we note that the 
more realistic case when the energy of two-body depends on $b_\omega$ and $D$ this effect is smoothed when studying the energy of the 
trimer with respect of the two-body energy.

\section{Conclusions}

In this article we used the Born-Oppenheimer approximation to study a 
heavy-heavy-light system. We derived the effective heavy-heavy potential for $D$ 
dimensions that allows the appearance of the Efimov effect. The effective potential 
gives also the dimensional interval where the Efimov effect is possible. Our results 
generalize the pioneering work of Fonseca and collaborators \cite{fonseca} 
that deduced long ago the form of the effective long-range
potential close to the unitary limit in three-dimensions. The present effective 
potential reproduces the well-known results for two and three dimensions. We applied the 
present analysis to the $^{133}$Cs-$^{133}$Cs-$^6$Li system and computed the scaling 
function obtained as a limit cycle of the correlation between the energies of two 
successive Efimov states with dependence on the heavy-light binding energy and dimension. 
We found that for $\sqrt{E^{D}_2/E_3^{(N)}}=0.89$ the $(N+1)$th excited state reach the 
two-body continuum, curiously, independent of the dimension. 
For D=3, Refs. \cite{wangprl2012,blumeprl2014} have studied the Efimov physics of the 
four-body mass-asymmetric systems. However, it still remains as a challenge the study 
of dimensionality reduction for bound mass-asymmetric systems beyond three bodies.

\ack
This work was partly supported by funds provided by the Brazilian agencies 
Funda\c{c}\~{a}o de Amparo \`{a} Pesquisa do Estado de S\~{a}o Paulo - FAPESP grants no. 
2017/05660-0 (TF), 2016/01816-2(MTY)  and 2013/01907-0(GK), Conselho Nacional de Desenvolvimento 
Cient\'{i}fico e Tecnol\'{o}gico - CNPq grants no. 305894/2009(GK), 142029/2017-3(DSR), 
302075/2016-0(MTY) and 308486/2015-3 (TF), Coordena\c{c}\~{a}o de Aperfei\c{c}oamento de 
Pessoal de N\'{i}vel Superior - 
CAPES no. 88881.030363/2013-01(MTY).

\end{document}